# Theoretical analysis of STM measurements

Kamil Walczak [1]

Institute of Physics, Adam Mickiewicz University
Umultowska 85, 61-614 Poznań, Poland

Modeling of electron transport through organic molecules is presented in order to interpret the experimental data of Rosink *et al*. [PRB **62**, 10459 (2000)]. Such results are understand as coherent off-resonance tunneling through the junction composed of molecular wires weakly coupled to the electrodes (Au-substrate and STM-tip, respectively). The influence of physical conditions on the characteristic parameters of the model is discussed in detail.

Electron transport through self-assembled organic molecules has been widely studied in recent years, using scanning tunneling microscope (STM) [1-3]. The STM allows electrical conduction of one or a few molecules at a time. The measurements we are trying to interpret are off-resonance with respect to the molecular orbitals, i.e. electrons tunnel at energies in the HOMO-LUMO gap (distance from the Fermi level to both mentioned energies is large enough to avoid resonant tunneling). In this situation (off-resonance limit) it is possible to treat the molecule as a tunneling barrier between two metal electrodes and simultaneously describe transport process in a simple Wentzel-Kramers-Brillouin (WKB) approximation [1,4-7]. In such a simplified picture, positions of the HOMO or LUMO levels determine the height of the tunnel barriers and charge transfer from metal to molecule have influence on its shape [6]. That advantageous formulation is based on a limited number of physical aspects (molecule length, effective barrier), what makes the experimental system more transparent. However, discussed model ignores the electronic structure of the molecule, although molecule itself is responsible for the observed behavior of current-voltage (I-V) curve. Furthermore, the conclusion of penetrating molecular layer by the STM tip for longer molecules may be speculative [1,8].

In this work we propose an alternative approach to theoretical considerations of STM measurements. Our parametric model incorporates molecular structure into the electron transfer scheme as well as idealized description of the contacts, considering off-resonance tunneling of electrons across a single molecule in response to the applied voltage. In order to interpret experimental results, we have used the data published by Rosink *et al*. [1], but anyway a brief summary of their experimental work is desired. Organic molecules were grown as self-assembled monolayer films, by selective chemisorption process. The monomers are benzene derivatives with two functional groups (–SH, –NH$_2$ or –CHO) in the para position: 4-aminothiophenol (4-ATP), terephthaldicarboxaldehyde (AT) and 1,4-diaminobenzene (ATD) – as shown in Fig.1. The aromatic rings are linked together by carbon-nitrogen double bonds (the imine bonds) and π conjugation extends over all atoms, whereas the coupling with the sulfur is reduced [8]. Such reduction is one possible factor that can affect the effective coupling parameter, as will be discussed later in this paper. I-V spectra were taken for different feedback voltage and current setpoints (i.e. for different initial sample-tip distances). All the experimental results represent the average of a subset from the collected I-V curves, where only those were included in the averaging that were going through the setpoint (details are included in [1]).

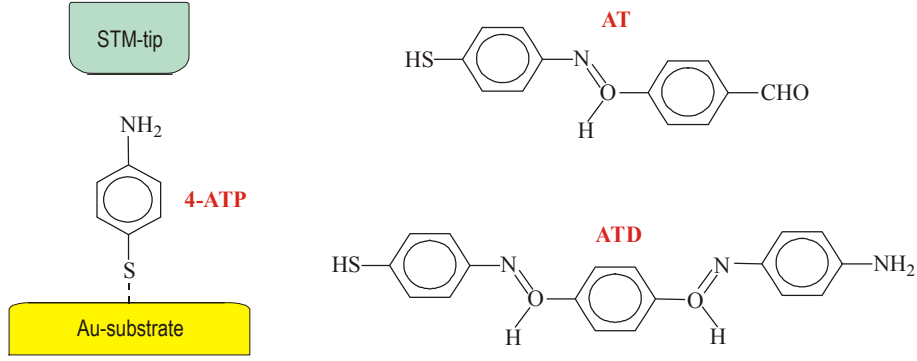

Fig.1 Schematic picture of the molecular junction and molecules under investigation.

At low bias, it is expected that one orbital will dominate for the transmission (HOMO or LUMO in this case) whichever is closest to the Fermi level, although the special extension of the orbitals might have some influence on the final results. Transmission probability through a single molecular level is well described by the Breit-Wigner formula [9-10]:

$$T(\varepsilon) = \frac{\Gamma_1 \Gamma_2}{[\varepsilon - \varepsilon_0]^2 + \Gamma^2}, \qquad (1)$$

where: $\Gamma = (\Gamma_1 + \Gamma_2)/2$, $\Gamma_1$ and $\Gamma_2$ are the so-called width parameters regarded to the coupling with the electrodes (or broadening of particular energy levels), $\varepsilon_0$ is the energy of the molecular orbital available for resonant tunneling. The width of the molecular level is exponentially sensitive to the distance from the electrode. The estimation associated with golden rule gives: $\Gamma_{1/2} \sim t_{1/2}^2 / D_{1/2}$, $t_{1/2}$ being an effective coupling parameter and $D_{1/2}$ being the energy bandwidth of the electron reservoirs (typical bandwidth of the metal electrodes is $D = 10$ eV). The tunneling current is derived according to the Landauer formula written in the form with a voltage division factor [3,11,12]:

$$I(V) = \frac{2e}{h} \int_{\varepsilon_F - \eta eV}^{\varepsilon_F + [1-\eta]eV} T(\varepsilon) d\varepsilon. \qquad (2)$$

Here, we also take a zero-temperature limit, which is allowed because the thermal energy $k_B T \sim 25$ meV is insignificant in comparison to the applied bias of 1.5 eV.

Inserting (1) into (2) allows us to obtain analytic formula for the current flowing through the device:

$$I(V) = \frac{2e}{h} \frac{\Gamma_1 \Gamma_2}{\Gamma} \left[ \arctan\left(\frac{\Delta + [1-\eta]eV}{\Gamma}\right) - \arctan\left(\frac{\Delta - \eta eV}{\Gamma}\right) \right], \qquad (3)$$

where: $\Delta = \varepsilon_F - \varepsilon_0$. Electrostatics of the problem is included as an effect of a voltage division factor $\eta$, which is directly proportional to the relative strength of the coupling with the electrodes [10]: $\eta = \Gamma_1/(\Gamma_1 + \Gamma_2)$. It means that voltage drop is assumed to occur only at the molecule/electrode interfaces, while the energy spectrum of the molecule remains unperturbed by applied bias (and hence the transmission function is treated as voltage-independent). The voltage division parameter is useful to describe the potential profile for short molecules, where the electric field inside the molecule seems to have a minimal effect on transport characteristics [3]. However, in the case of longer molecular wires, the electric field inside the molecule may play a more significant role depending on the internal structure of the molecule. Also at higher voltages, changes in the charge density inside the molecule could be significant and its contribution to the potential profile would lead to effective changes in the value of $\eta$.

TABLE I. Parameters resulting from fits to the presented model
($\Gamma_1$ and $\Gamma_2$ are given in $10^{-2}$ eV, $\Delta$ is given in eV).

| Sample | 500 mV, 100 pA | | | 500 mV, 200 pA | | | 500 mA, 500 pA | | |
|---|---|---|---|---|---|---|---|---|---|
| | $\Gamma_1$ | $\Gamma_2$ | $\Delta$ | $\Gamma_1$ | $\Gamma_2$ | $\Delta$ | $\Gamma_1$ | $\Gamma_2$ | $\Delta$ |
| 4-ATP | 0.15 | 0.20 | 1.047 | 0.21 | 0.26 | 0.932 | 0.25 | 0.35 | 0.872 |
| AT | 0.22 | 0.24 | 1.016 | 0.31 | 0.32 | 0.968 | 0.37 | 0.37 | 0.805 |
| ATD | 0.14 | 0.18 | 1.107 | 0.24 | 0.34 | 1.017 | 0.26 | 0.39 | 0.780 |

The agreement between our theoretical analysis and observed I-V characteristics is obtained by fitting parameters of our model over reasonable limits (see Fig.2). All the fit results are listed in Table I. It should be noted general tendencies: the shorter molecule-electrode distance, the greater values of the width parameters and the smaller values of $\Delta$. One surprising aspect of our results that requires comment is associated with the actual values of the width parameters ($\Gamma_1$ and $\Gamma_2$). Such quantities depend on the local DOS of the metal surface (unstructured by assumption) and the character of the molecule-to-electrodes coupling, while the voltage division factor takes into account asymmetries due to the nature of the contacts. To indicate that basic problem we have to investigate the chemical conditions of the junction formation. When thiol-terminated molecule is brought close to a gold substrate, the end hydrogen atom of the end group is lost in the chemisorption process and the sulfur atom bonds with three gold atoms arranged in an equilateral triangle [13]. For a good contact EHT predicts the effective coupling parameter between sulfur $p_z$ orbital and three gold s orbitals of about t = 2 eV in the direction parallel to the electrode surfaces [14].

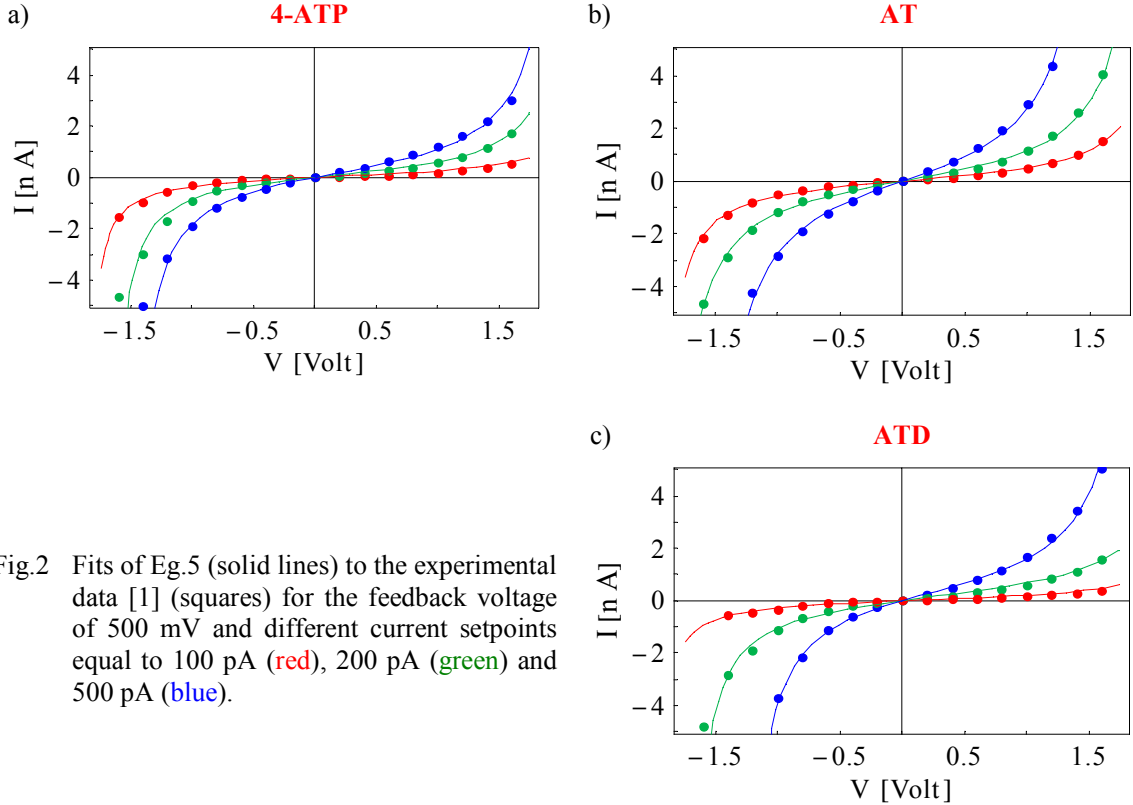

Fig.2 Fits of Eg.5 (solid lines) to the experimental data [1] (squares) for the feedback voltage of 500 mV and different current setpoints equal to 100 pA (red), 200 pA (green) and 500 pA (blue).

It suggests that $\Gamma = 0.40$ eV. However, in order to reproduce experimental values of the current, it is assumed the width parameters to be two orders of magnitude smaller. The origin of such discrepancy can be attributed to the contact geometry and also to its chemistry [15].

Furthermore, since sulfur is more electronegative than gold, so the coupling between them should be a polar one and Schottky barrier could occur for moving electrons. Since Au-S distance is about 1.9 Å and typical charge transfer is expected to be $-0.2e$ to $-0.4e$ [13], the Au-S dipole moment is deduced to be 0.9D to 1.8D [1]. It is difficult to estimate what it means for electronic transport phenomena. However, our results indicate that voltage division factor is $0.4 < \eta \leq 0.5$ (as suggested in [3]), so contribution of the Au-S dipole is negligibly small because of the screening effects [16,17].

Additional factors that can alter the value of the current are associated with temperature effects (hot electrons and vibrational coupling) or local disorder in the electrode near the contacts (electron localization) [15]. Anyway, the shape of the I-V curve is determined by the electronic structure of the molecule in contact with the electrodes and in the presence of an external electric field.